\documentclass[namedreferences]{solarphysics}
\usepackage[hyperref,optionalrh,solaromanenum]{spr-sola-addons} 
\usepackage{graphicx}                    
\usepackage{color}                       
\usepackage{breakurl}                         

\chardef\us=`\_

\begin{document}

\begin{article}

\begin{opening}

\title{Observations of Ray-Like Structures in Large-Scale Coronal Dimmings Produced by Limb CMEs}

%
\author[addressref=aff1,corref,email={goryaev\underline{ }farid@mail.ru}]{\inits{F.F.}\fnm{F.F.}~\lnm{Goryaev}\orcid{https://orcid.org/0000-0001-9257-4850}}
\author[addressref=aff1]{\inits{V.A.}\fnm{V.A.}~\lnm{Slemzin}\orcid{https://orcid.org/0000-0002-5634-3024}}
\author[addressref=aff1]{\inits{D.G.}\fnm{D.G.}~\lnm{Rodkin}\orcid{https://orcid.org/0000-0002-5874-4737}}

%
\runningauthor{F.F.~Goryaev {\it et al.}}
\runningtitle{Observations of Ray-Like Structures in Large-Scale Coronal Dimmings}

\address[id=aff1]{P.N.~Lebedev Physical Institute RAS, 53 Leninskiy Prospect, Moscow, 119991, Russia}

\begin{abstract}
Observations of the off-limb corona with the {\it Sun Watcher with Active Pixels and Image Processing} (SWAP) wide-field telescope in the 174\,\AA\ passband onboard the {\it Project for Onboard Autonomy 2} (PROBA2) mission provide an opportunity to study post-eruptive processes in the dimming regions. We investigate morphology, temporal evolution, and plasma properties for four `deep' off-disk coronal dimmings associated with limb coronal mass ejections (CMEs) in 2010\,--\,2017. Using the SWAP fixed-difference images, we revealed ray-like structures that appeared in the dimming recovery phase stretching quasi-radially to distances from 1.1 to 1.6\,$\mathrm{R}_{\odot}$ and existing from tens of minutes to several hours. Similar rays were detected earlier at distances above 1.7\,$\mathrm{R}_{\odot}$ by the {\it Ultra-Violet Coronagraph Spectrometer} (UVCS) onboard the {\it SOlar and Heliospheric Observatory} (SOHO). These structures apparently represent the coronal roots of the flux rope trunks observed by the {\it Large Angle Spectroscopic COronagraph} (LASCO) C2 onboard SOHO. The {\it Extreme-Ultra-Violet Imager} (EUVI) data onboard the {\it Solar TErrestrial RElations Observatory} (STEREO) show the origins of these structures on the disc as fan rays much brighter in the 171\,\AA\ band than in 193\,\AA , which suggests their temperature being less than 2\,MK. The differential emission measure (DEM) analysis based on the {\it Atmospheric Imaging Assembly} (AIA) multi-wavelength images onboard the {\it Solar Dynamics Observatory} (SDO) showed that the emission measure (EM) in these rays compared to the pre-eruption plasma state increased up to 45\,\% at temperatures of 0.6\,--\,0.8\,MK, whereas EM of the ambient coronal plasma with temperatures of 1.3\,--\,3.7\,MK dropped by 19\,--\,43\,\%. For the event on 18 August 2010, the {\it PLAsma and SupraThermal Ion Composition} (PLASTIC) instrument onboard STEREO detected signatures of the cold streams in the CME tail as enriched with the ions Fe$^{8+}$\,--\,Fe$^{10+}$, which may be associated with the post-eruptive rays in the solar corona.
\end{abstract}

%

\end{opening}

%

\section{Introduction}
     \label{S-Introduction}

Coronal mass ejections (CMEs) are large-scale manifestations of solar activity, often followed by transient dark regions known as coronal dimmings. Dimmings (``coronal depletions'') were originally identified by \cite{Hansen74} in white light with the Mauna Loa ground-based coronagraph and by \cite{Rust76} in the soft X-ray {\it Skylab} images. These temporal depletions appear as dark areas similar to coronal holes and thus they are sometimes referred to as ``transient coronal holes'' \citep{Rust83}. \cite{Sterling97} investigated a dimming on the solar disc associated with a halo CME, observed by the {\it Soft X-ray Telescope} (SXT) onboard {\it Yohkoh}. \cite{Thompson98} considered dimmings as markers of the eruption regions at the solar disc observed by the {\it Extreme-Ultraviolet Imaging Telescope} (EIT) onboard the {\it SOlar and Heliospheric Observatory} (SOHO) in the 195\,\AA\ spectral band. They suggested that the decrease in brightness in the dimming occurs due to a decrease in plasma density as contrasted to the decrease due to change in temperature. This conclusion was later confirmed by \cite{Zarro99} in observations of a coronal dimming associated with a halo CME by SOHO/EIT simultaneously with {\it Yohkoh}/SXT. The appearance of extreme-ultraviolet (EUV) and soft X-ray (SXR) coronal dimmings were interpreted within the framework of a flux rope eruption, partially controlled by the CME.  In a spectroscopic study with the {\it Coronal Diagnostic Spectrometer} (CDS) on board SOHO, \cite{Harrison} have found activation of an adjacent prominence and 2~MK ``hot spots'' under the ascending CME, which may be associated with the CME footpoints. Using the method of differential imaging to observe dimmings in the EUV lines, \cite{Chertok03} found coincidence of long-lived darkenings on the solar disc in different EUV bands of SOHO/EIT corresponding to coronal temperatures and in the 304\,\AA\ transition-region band. However, they found distinctions between images of dimmings in different emission lines, which suggest that the coronal plasma inside the dimmings is variable in temperature. \cite{Aschwanden06} summarized the results of the dimming investigation in the SOHO era: they are detected a relative deficit of coronal mass or emission measure compared with pre-CME conditions, interpreted as a vacuum-like rarefaction or ``evacuation'' of the coronal plasma after the launch of a CME. The appearance of coronal dimmings is most dramatically seen on the solar disk, but they are also detectable above the solar limb in favorable cases, when an eruptive active region is located near the limb. Dimmings are regarded as a relevant constituent of the CME evolution and as a probable source of solar wind.

\cite{Harrison03} were the first to describe the plasma diagnostic characteristics of a set of EUV dimming events on the solar limb associated with CME onsets. They found that the dimming events developed synchronously with the projected CME onset and spatially coincided with the CME origin locations of the associated CMEs. \cite{Zhukov04} investigated several examples of EIT waves, EUV coronal dimming regions and strengthened their link to CMEs. They performed a differential emission measure (DEM) analysis and found that the mass of about 10$^{15}$\,g erupts from the dimming region. It appeared that around 50\,\% of total CME mass was initially contained in the low corona outside of the ``transient coronal holes''.

\cite{Bein13} and \cite{Temmer17} studied a mass evolution in the outer corona for a set of CMEs. They found a distinct mass increase in propagating CMEs produced by mass flow from the corona behind them. \cite{Mandrini07} analyzed magnetic structure of the dimming regions and proposed the scenario of dimming formation as a combination of core and secondary dimmings originated from reconnection of the surrounding coronal magnetic loops. \cite{Vanninathan18} applied DEM diagnostics to study the plasma characteristics of coronal-dimming events. They found that secondary dimmings are less deep and have less duration in comparison with the core dimmings. In a recent work, \cite{Veronig19} studied a coronal dimming caused by a fast halo CME using the {\it Extreme-ultra-violet Imaging Spectrometer} (EIS) spectroscopy onboard {\it Hinode} and the {\it Atmospheric Imaging Assembly} (AIA) onboard the {\it Solar Dynamics Observatory} (SDO) DEM analysis. They investigated temporal dynamics of the relative density changes in the core dimming regions and concluded that the CME mass increase in the coronagraphic field of view results from plasma flows from the lower corona derived from the EIS data.

White-light and EUV coronagraphic observations of the solar corona often reveal long-lived, enhanced ray-like structures associated with streamers or pseudo-streamers associated with heliospheric current sheets (see, {\it e.g.}, \citealp{Wang00}; \citealp{Wang07}; \citealp{Slemzin08}, and references therein). \cite{Wang07} estimated the flow speed from pseudo-streamers to be about 200\,km\,s$^{-1}$ at heliocentric distances of $\approx 3\,\mathrm{R}_{\odot}$ supporting the prediction (based on their low flux tube divergence rates) that pseudo-streamers are sources of fast solar wind. \cite{Crooker14} found that solar-wind flows from pseudo-streamers had similar characteristics to the streamer flows (low speed and proton temperature, high density and composition ratio), however, these characteristics in pseudo-streamers were slightly less pronounced.

\cite{Munro85} reported {\it Skylab} observations of transient brightenings at the base of pre-existing streamers and ray-like features, which they identified as a special class of coronal transients (CMEs). Ray-like coronal features formed in the aftermath of CMEs were studied by \cite{Webb03}, \cite{Bemporad06}, \cite{Vrsnak09}, \cite{Ciaravella08,Ciaravella13} as signatures of  large-scale current sheets created by magnetic reconnection. \cite{Ciaravella13} distinguished hot, coronal, and cool rays that appeared in the aftermath of CMEs and found that about 18\,\% of the white-light rays show very hot gas consistent with the current sheet interpretation, while about 23\,\% contained cold gas that they attributed to cool prominence material draining back from the CME core. Physical conditions of the CME plasma after eruption, its heating, energy balance, and mass evolution were studied using the {\it Ultra-Violet Coronagraph Spectrometer} (UVCS) onboard SOHO and Hinode/EIS spectroscopic data by \cite{Bemporad07} and \cite{Lee09}. \cite{Landi10,Landi13} explored a hot plasma associated with the CME of 9 April 2008 using the {\it Hinode}/EIS and XRT data at a distance $1.1\,\mathrm{R}_{\odot}$.

However, due to the gap between the observational data obtained with coronagraphs at distances exceeding 2\,$\mathrm{R}_{\odot}$ and the data of most EUV telescopes obtained below 1.3\,$\mathrm{R}_{\odot}$, the phenomena in the corona accompanying CMEs have not been studied in the full range. The launch of the {\it PRoject for OnBoard Autonomy 2} (PROBA2) spacecraft with the {\it Sun Watcher using Active Pixel System Detector and Image Processing} (SWAP) wide-field EUV telescope in 2009 made possible systematic studies in the EUV 174\,\AA\ band coronal sources of the structures observed at larger distances by white-light coronagraphs. The main advantages of SWAP are a wide field of view (up to 1.7\,$\mathrm{R}_{\odot}$ from the center of the Sun and up to 2.2\,$\mathrm{R}_{\odot}$ in the off-point positions) and very low stray-light permitting EUV observations of the corona at the distances inaccessible for the most of EUV telescopes. \cite{Goryaev14} studied properties of a streamer in the beginning of Solar Cycle 24 combining the data of PROBA2/SWAP, {\it Hinode}/EIS, and Mauna Loa Mk4 instruments. They found that the plasma density in the streamer as a function of height corresponded to scale height temperature higher than the excitation temperature, which suggests the presence of plasma outflow along the streamer. SWAP observations of the active events at the limb disclose large-scale restructuring of the corona; in particular, appearance of dimmings and bright coronal rays accompanying the CME event. Such measurements are important to explore primary solar-wind sources significant for the space-weather applications.

The aim of the current work is to study bright ray-like structures observed by SWAP in the dimming regions and to analyze their properties in the corona and their links to solar wind using the capabilities of the SWAP and SDO/AIA telescopes and special methods of data analysis. We investigated morphology, temporal evolution, and plasma properties of several deep coronal off-disk dimmings observed by SWAP in 2010\,--\,2017. To study the plasma characteristics of the coronal dimmings, we used DEM analysis based on the multi-wavelength SDO/AIA EUV data. The correspondence of these dimmings to specific solar-wind features was studied using data of the {\it PLAsma and SupraThermal Ion Composition} (PLASTIC) instrument onboard the {\it Solar TErrestrial RElations Observatory} (STEREO) Ahead (STEREO-A) and Behind (STEREO-B) spacecrafts. The article is organized as follow. Observations of the coronal structures under study by SWAP and other instruments, as well as data processing, are described in Section 2. In Section 3, we describe the DEM analysis of the plasma parameters. Discussion of the results is given in Section 4. Finally, Section 5 presents the summary and conclusions. Electronic Supplementary Material (movies) illustrates development of the ray-like structures in the dimmings in all four events.

\section{Observations and Data Processing}

Table~\ref{T-1} lists some of the observational information on the four events with the deepest dimmings taken from the Solar Demon database (\citealp{Kraaikamp15}; online database {\sf solardemon.oma.be/}).

\begin{table}
\caption{Coronal-dimming events under study (data from {\sf solardemon.oma.be/}). All times are in UT.
}
\label{T-1}
\begin{tabular}{clrccccrc}     
  \hline                   
Event & Date & Min. & Start & Peak & End  & AR & Position & Flare \\
     &      & int.\tabnote{Values of the integrated decrease of intensity over the dimming area during the observation period from the AIA 211\,\AA\ images.} & time  & time & time &    & Angle\tabnote{Position angles are indicated clockwise from the North pole (see Figure~1a).}    &    \\
  \hline
1 & 18 Aug. 2010 & -1116  & 05:02 & 06:00 & 07:02 & 4359 & 70 & C7 \\
2 & 08 Mar. 2011 & -1104  & 03:42 & 03:48 & 04:46 & 4485 & 250 & M2 \\
3 & 27 Jan. 2012 & -1795  & 18:06 & 18:56 & 20:08 & 5333 & 60 & X3 \\
4 & 10 Sep. 2017 & -945  & 15:51 & 16:17 & 17:29 & 9042 & 100 & X3 \\
  \hline
\end{tabular}
\end{table}

The SWAP EUV telescope launched onboard the PROBA2 microsatellite on 2 November 2009 is capable of imaging the solar disk and inner corona in a single spectral band, 174\,\AA , over a 54$\times$54~arcmin field of view with 3.17 arcsec resolution and a cadence of about two minutes (see \citealp{Seaton13} for details of the design and parameters of SWAP). The response function of SWAP covers the temperature range approximately from 0.5 to 2\,MK (\citealp{Raftery13}), which embraces four of the most intense spectral lines of Fe ions: Fe~{\sc ix} $\lambda$171.08, Fe~{\sc x} $\lambda$174.53, Fe~{\sc x} $\lambda$177.24, and Fe~{\sc xi} $\lambda$180.41. The telescope has a two-mirror optical design including special baffles to suppress stray light over the field of view up to the heliocentric distance 1.7\,$\mathrm{R}_{\odot}$ to the level of 10$^{-3}$ of the disc intensity without correction and to $4\times 10^{-5}$ with the point spread function (PSF) correction (\citealp{Goryaev14}).

  \begin{figure}    
   \centerline{
               \includegraphics[width=0.340\textwidth,clip=]{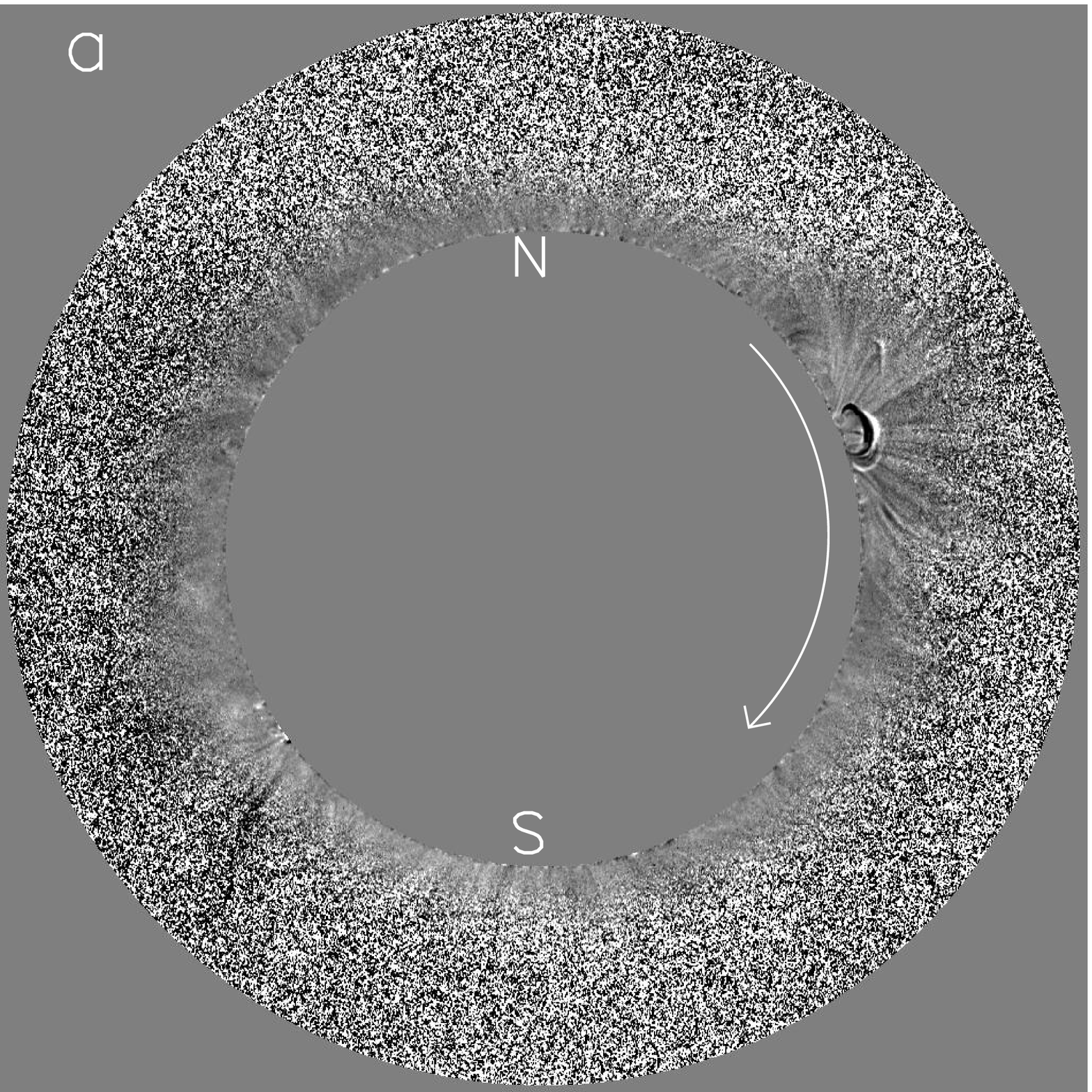}
               \includegraphics[width=0.6\textwidth,clip=]{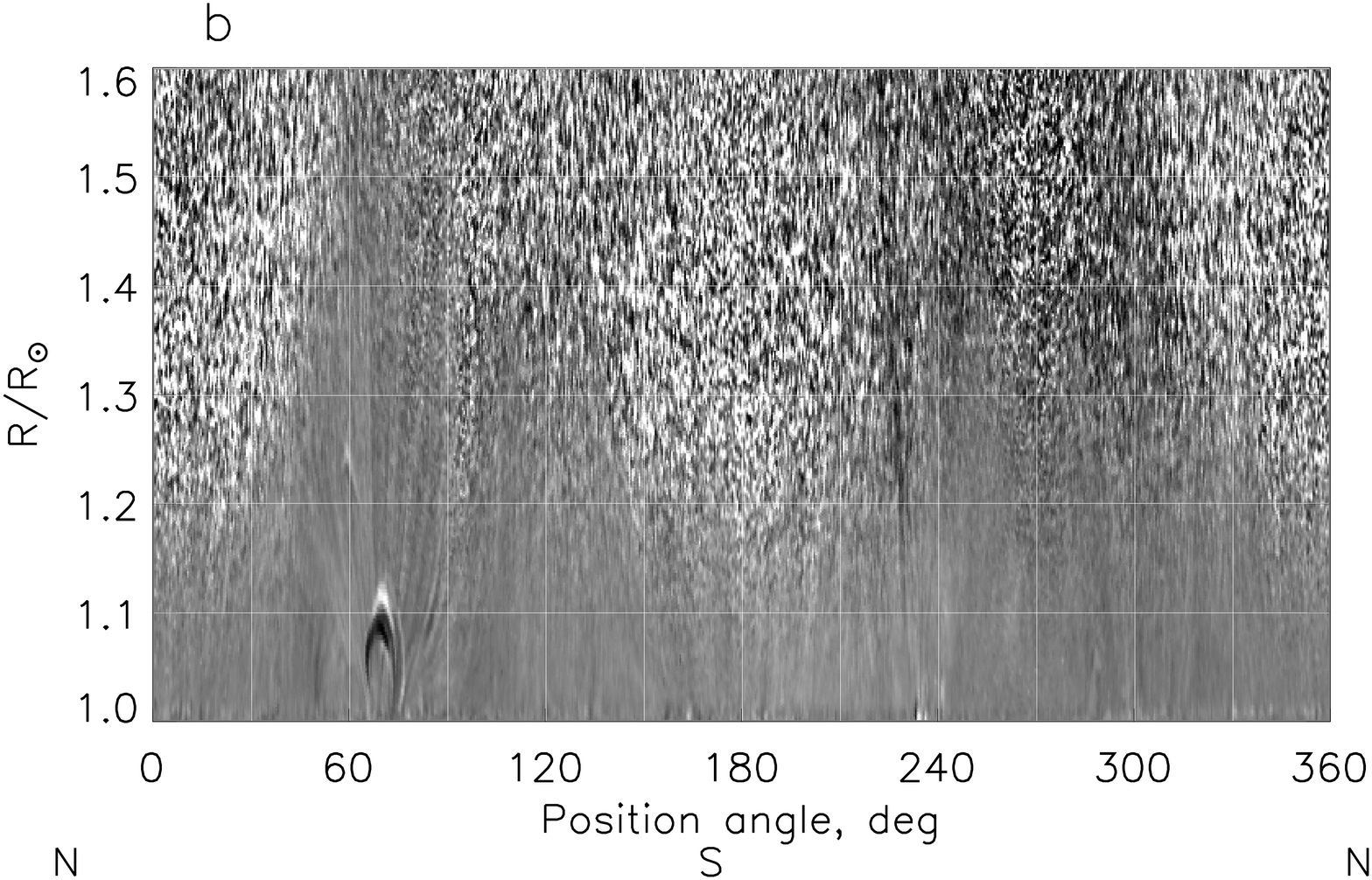}
              }
\caption{SWAP relative difference images on 18 August 2010 (04:30:17\,--\,04:00:17\,UT): (a) initial disc image, (b) its representation in polar coordinates. The arrow shows the direction of increase of position angle.
        }
   \label{F-1}
   \end{figure}

The procedure of processing the SWAP images consisted in converting SWAP level 0 FITS files into level 1 using the standard {\sf p2sw\_prep} Solarsoft procedure, which includes correction for dark current, detector bias, flat-field variations, and bad pixels with the option to apply a PSF deconvolution to the data for removing stray light (see \citealp{Halain13}). The effectiveness of this procedure was also estimated by \cite{Goryaev14}, where the level of stray-light after correction constituted less than 20\,\% of the coronal emission up to the heliocentric distances of 1.7\,$\mathrm{R}_{\odot}$.

To emphasize the ray-like coronal structures in the dimming regions, the SWAP images of the off-limb corona were then transformed to polar coordinates using an interpolation procedure with taking into account the difference of the pixel dimensions in both representations. The format of these polar images was of 720$\times$400 pixels with resolution of 0.5 degrees in latitude and $1.5\times 10^{-3} \,\mathrm{R}_{\odot}$ in height. To observe variations of the dimming structure brightness at large heights where the intensity of light was seriously decreased, we analyzed the relative difference images (present image $I(t)$ minus reference image $I(t_0)$ divided by the reference image $I_{\mathrm{ref}}(t_0)$ excluding pixels with zero signal):

\begin{equation}
y(t) = \frac{I(t) - I_{\mathrm{ref}}(t_0)}{I_{\mathrm{ref}}(t_0)} \, .\label{Eq-1}
\end{equation}

\noindent Figure~\ref{F-1} shows an example of the SWAP relative difference image of the solar corona and its representation in polar coordinates for the event of 18 August 2010 before eruption. Such a representation allows to enhance extended radial coronal structures stretched outward from the solar surface.

  \begin{figure}    
   \centerline{\includegraphics[width=1.1\textwidth,clip=]{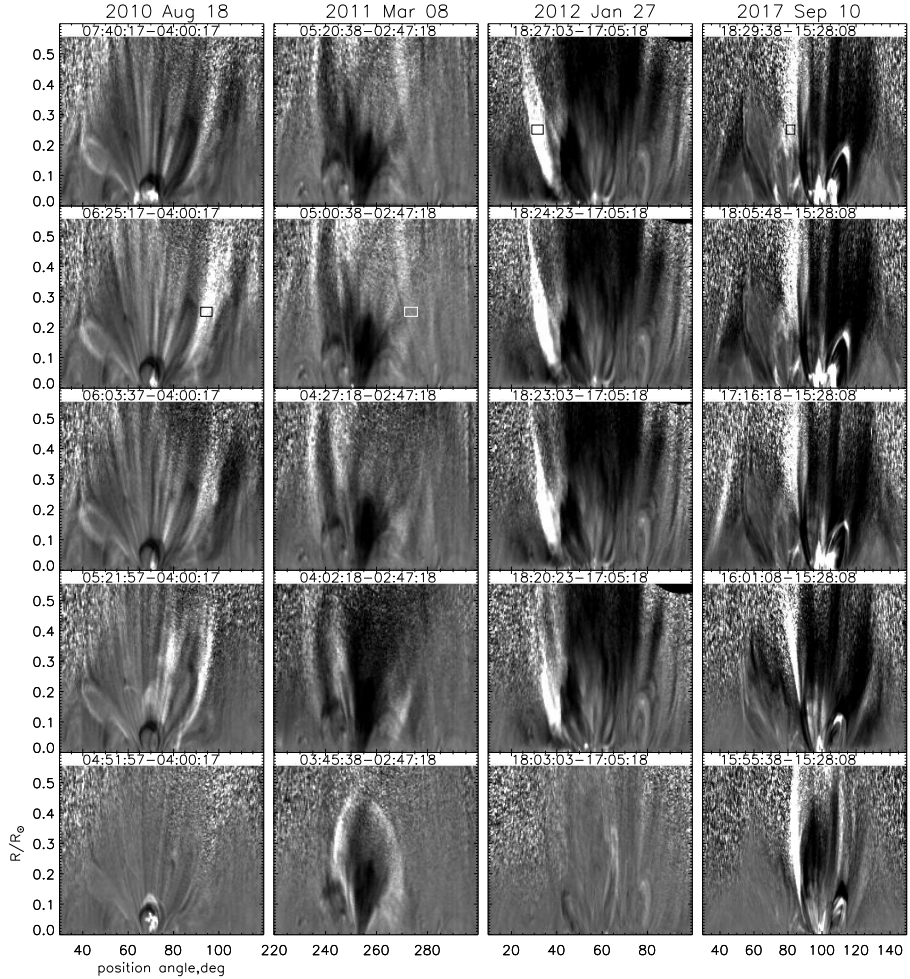}
              }
              \caption{Polar maps of the dimming regions at selected times for the events listed in Table~\ref{T-1}. The times of the given and reference images [UT] are shown in the titles of each frame. The brightness corresponds to relative difference of intensity in the SWAP images. The rectangles indicate the regions for the DEM analysis at the distance $1.25\,\mathrm{R}_{\odot}$.
                      }
   \label{F-2}
   \end{figure}

\begin{figure}    
   \centerline{\includegraphics[width=1.03\textwidth,clip=]{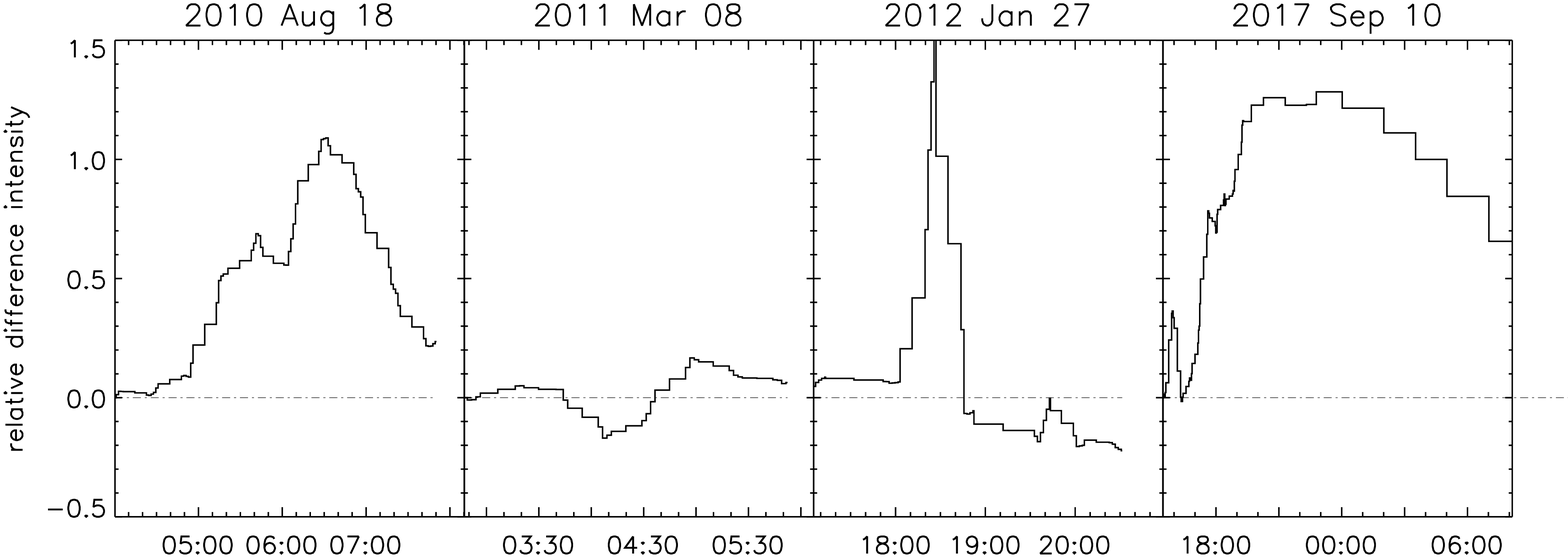}
              }
              \caption{Temporal profiles of the relative intensity in the selected regions of ray-like features at the SWAP difference images displayed in Figure~\ref{F-2}.
                      }
   \label{F-3}
   \end{figure}

The dimming regions were determined from these images by the condition $y<0$. Figure~\ref{F-2} presents the selected SWAP images of the dimming regions for the events listed in Table~\ref{T-1}. Development of the dimmings and appearance of the ray-like features in the events under study are illustrated in Electronic Supplementary Material with the movies in the polar and relative difference polar presentations. In all cases, in the recovery periods (later than the peak of the dimming areas) the dimmings contained distinct structural elements. Below heliocentric distances of about $1.2\, \mathrm{R}_{\odot}$, there are post-eruptive loops, whereas at greater heights quasi-radial ray-like structures are seen being brighter than the initial (before eruption) parts of the corona at the same place. Some of such rays appear in the traces of the expanding flux rope legs, and others appear some distance away from them. The rectangular boxes at the distance $1.25\,\mathrm{R}_{\odot}$ in some of the SWAP images in Figure~\ref{F-2} indicate the brightest places in the rays selected for the plasma diagnostics. The sizes of these regions are of 5.5$^{\circ}$ in latitude and 0.031\,$\mathrm{R}_{\odot}$ in height. Figure~\ref{F-3} shows temporal profiles of the relative intensity in the selected regions of the dimmings at the SWAP difference images.

  \begin{figure}    
   \centerline{\includegraphics[width=1.1\textwidth,clip=]{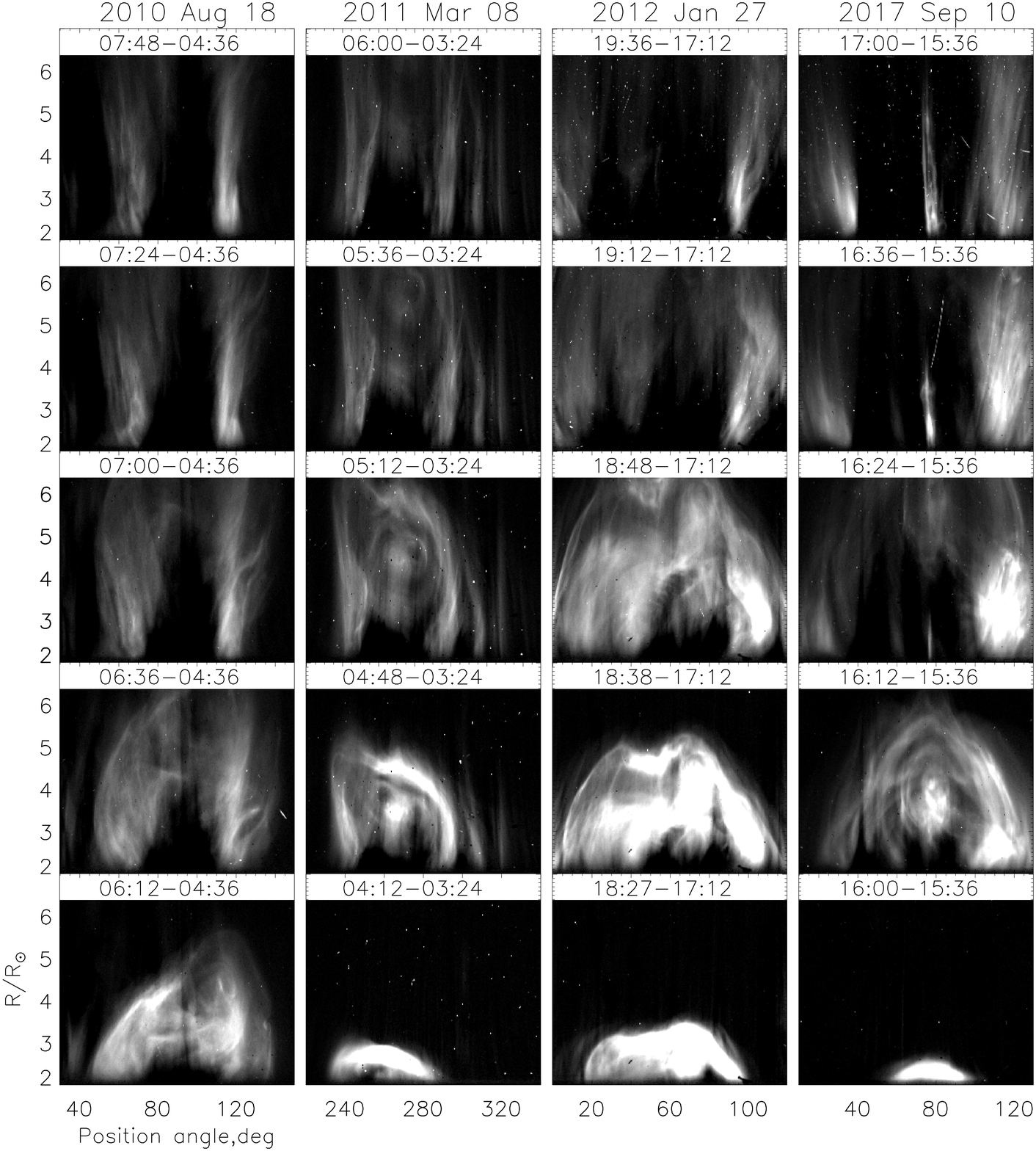}
              }
              \caption{The LASCO-C2 difference images (in polar representation) showing development of the CMEs under study at the distances 2.2\,--\,6\,$\mathrm{R}_{\odot}$.
                      }
   \label{F-4}
   \end{figure}

Figure~\ref{F-4} demonstrates propagation of the studied CMEs at the distances 2.2\,--\,6\,$\mathrm{R}_{\odot}$ seen by the {\it Large Angle Spectroscopic Coronagraph} (LASCO) C2 onboard SOHO (fixed-difference images in polar representation). These structures are evidently appeared as the lower parts (legs) of the flux ropes and some of them may be associated with extensions of the bright EUV features seen in the SWAP images in Figure~\ref{F-2}.

\begin{figure}    
   \centerline{
               \includegraphics[width=0.5\textwidth,clip=]{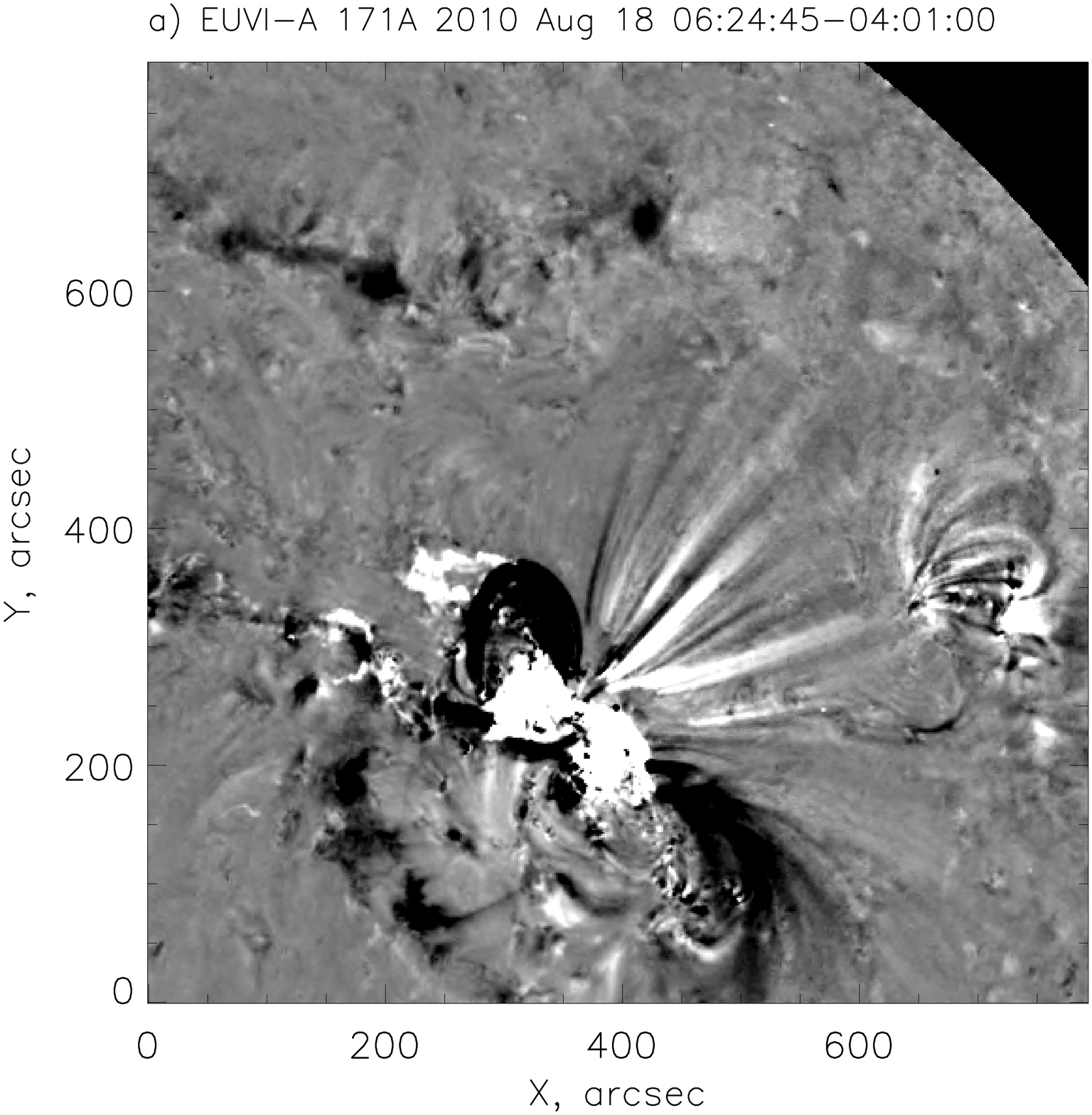}
               \includegraphics[width=0.5\textwidth,clip=]{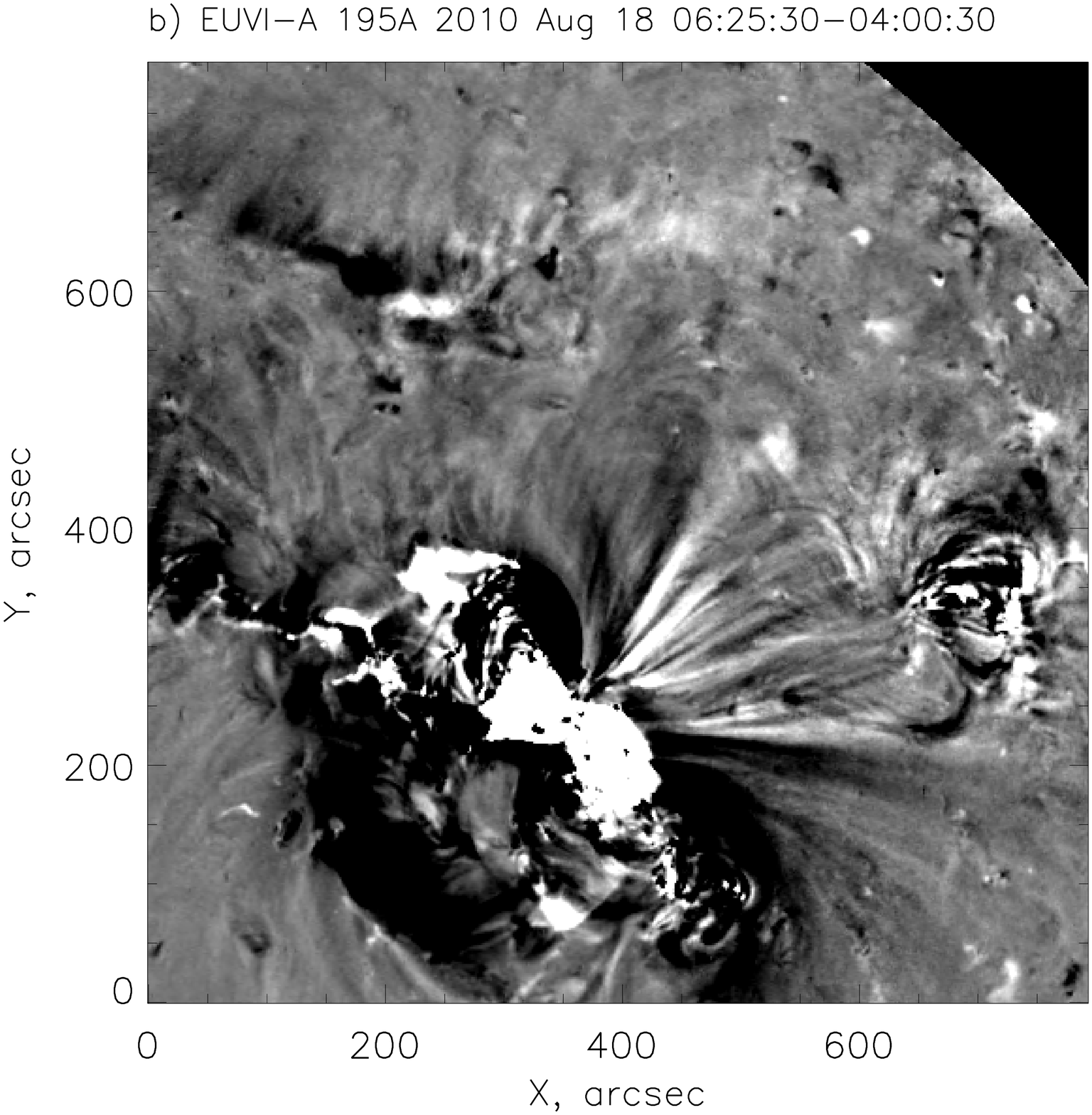}
              }
   \centerline{
               \includegraphics[width=0.5\textwidth,clip=]{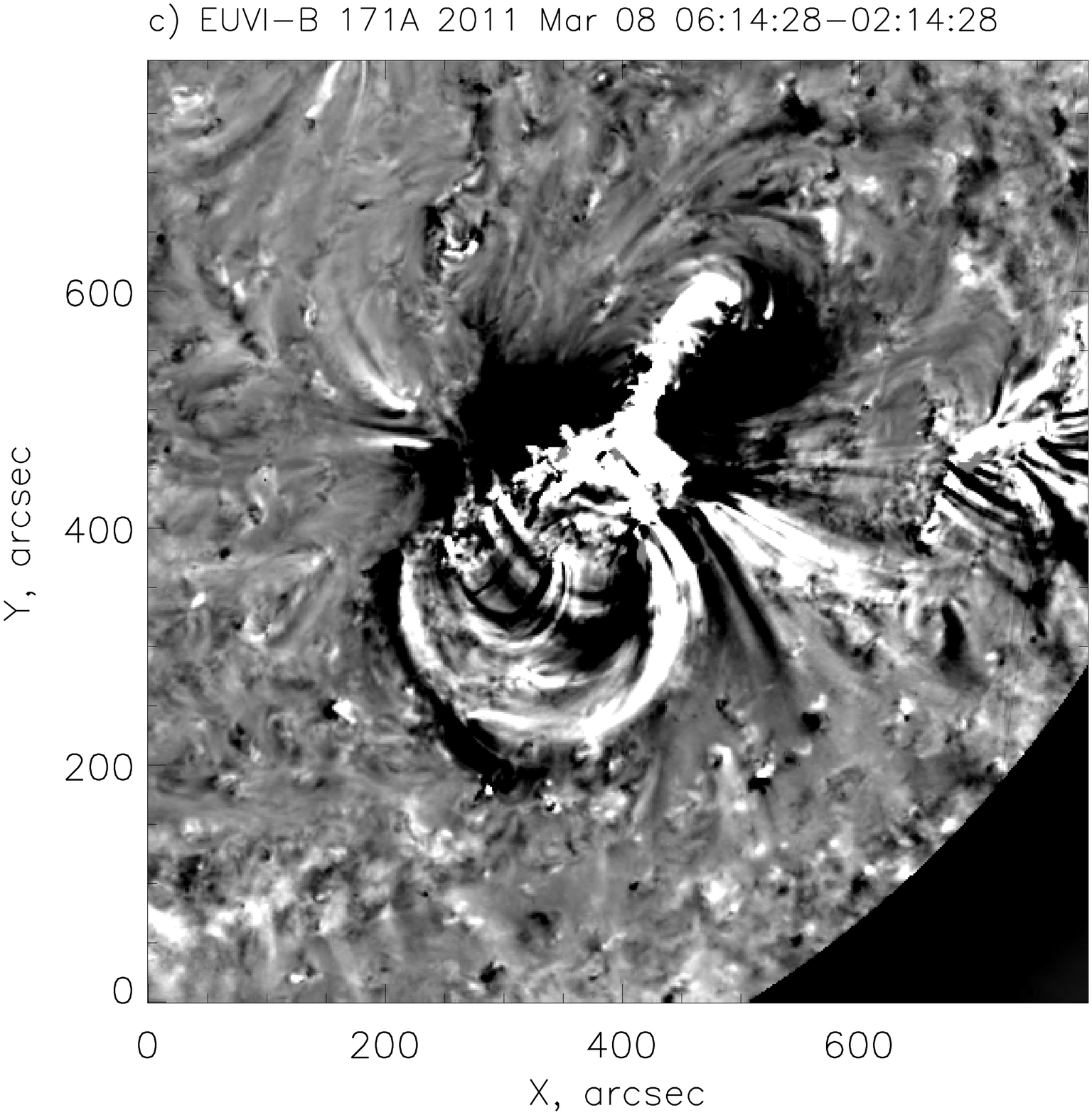}
               \includegraphics[width=0.5\textwidth,clip=]{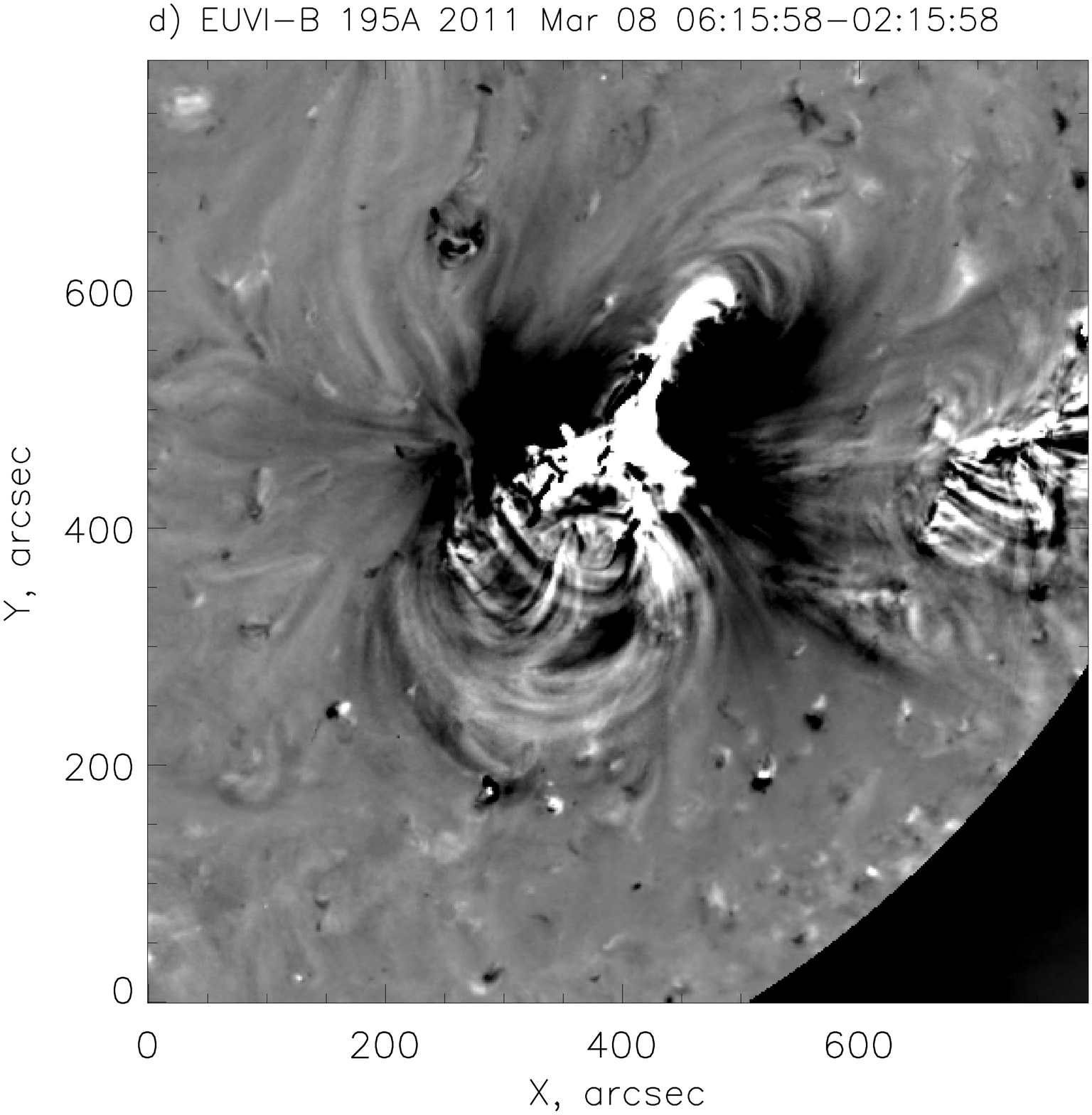}
              }
\caption{Top: STEREO-A/EUVI difference images of the dimming region in 171~\AA\ (a) and 195~\AA\ (b) wavelength channels for the event on 18 August 2010. Bottom: The same on 8 March 2011 (171\,\AA\ (c) and 195\,\AA\ (d)).
        }
   \label{F-5}
   \end{figure}

Figure~\ref{F-5} presents difference images of the dimmings on the disc derived from the {\it Extreme-Ultra-Violet Imager} (EUVI) data onboard STEREO-A (event on 18 August 2010) and STEREO-B (event on 8 March 2011). In the vicinity of active regions one can see the bright fan rays, which probably represent origins of the ray-like features seen by SWAP.  These fan rays in the 195\,\AA\ look dimmer than in the 171\,\AA\ band (or completely absent as in the second event on 8 March 2011), which suggests that their plasma has temperatures less than 2\,MK.

\section{DEM Analysis}

The DEM diagnostics of the coronal plasma in the dimming regions selected from the SWAP difference images was carried out using the SDO/AIA data in six EUV channels: 94, 131, 171, 193, 211, and 335\,\AA . To reconstruct the DEM temperature distributions, we used the fast code of \cite{Plowman13}. The AIA level~1 images were first corrected for PSF using the Solarsoft {\sf aia\_deconvolve\_richardsonlucy} routine followed by processing the level 1 images to the level 1.5 using the {\sf aia\_prep} code. The processed images were then binned from the original 4096$\times$4096 frames to frames of 1024$\times$1024 pixels to increase the signal-to-noise ratio (S/N) and to accelerate further treatment, and then they were used to retrieve the DEM temperature maps with the Plowman {\it et al.} code. For a better comparison with the SWAP data, the DEM maps were also transformed into polar coordinates with correction of values according to the geometric conjugation of the polar pixel area with the disc ones. For visual presentation, the derived DEM temperature distributions were divided into three temperature components transformed into three color channels of the true-color picture: the cool plasma with $\log_{10} T[\mathrm{K}] = 5.5-6.1$ ($T = 0.3 - 1.3$\,MK) is transformed into the blue component, the middle temperature plasma with $\log_{10} T[\mathrm{K}] = 6.1-6.5$ ($T=1.3 - 3.7$\,MK) into the green one, and the hot plasma with $\log_{10} T[\mathrm{K}] = 6.5-7.5$ ($T=3.7 - 30$\,MK) into the red one. The brightness in these channels is proportional to the emission measure in the corresponding temperature bands.

  \begin{figure}    
   \centerline{
               \includegraphics[width=1.0\textwidth,clip=]{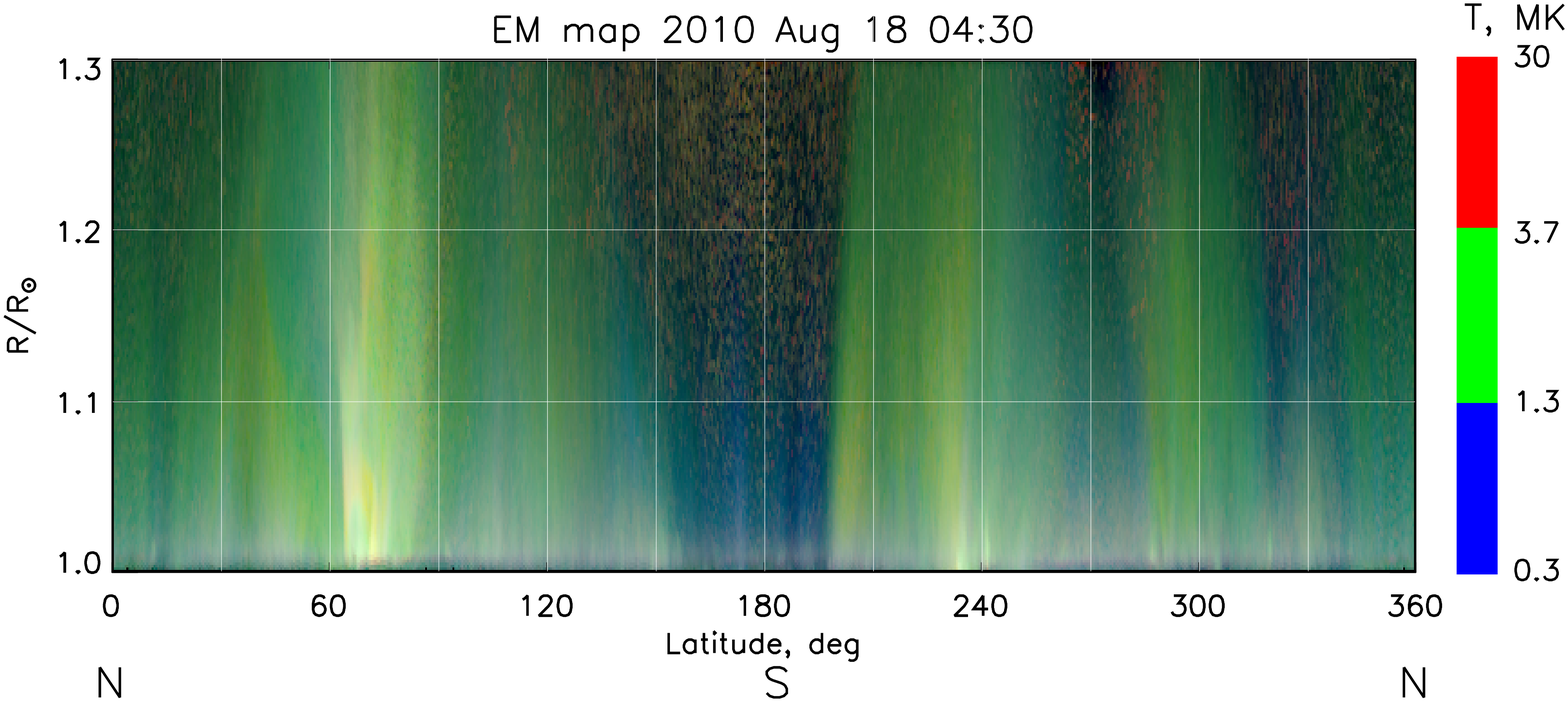}
              }
\caption{EM true color polar maps for the event 18 August 2010 (04:30~UT) for three temperature intervals: 0.3\,--\,1.3\,MK (blue), 1.3\,--\,3.7\,MK (green), and 3.7\,--\,30\,MK (red). Each point on the image is the mixture of these three colors. The position angle grows clockwise as in Figure~\ref{F-1}.
        }
   \label{F-6}
   \end{figure}

As it is seen from Figure~\ref{F-6}, such a color picture demonstrates clearly the distribution of the temperature components above the limb over the whole corona. The dominant green color enhanced in the equatorial regions shows the main unstructured component of the quiet corona. The red color ($T>3-4$\,MK) is concentrated in the hot loops above active regions in a mixture with other components. The blue color dominates in the polar regions where the temperatures constitute $T\lesssim 1$\,MK in the absence of other hotter components. The uncertainty of these DEM distributions derived by means of the Plowman {\it et al.} method is about 30\,--\,35\,\% in the emission measure (EM) values (that is proportional to $\approx N_e^2$). This value is summarized from 20\,\% calibration accuracy of different spectral bands, 10\,--\,15\,\% errors of the DEM reconstruction (revealed by comparison of the initial intensities with the recalculated ones using the obtained DEM) and $\approx$20\,\%  uncertainty for the atomic data from the CHIANTI database (\citealp{DelZanna18}). The peak temperature for each point of the DEM map was determined with the accuracy of $\log_{10}(T[\mathrm{K}])= \pm 0.1$, which corresponds to the typical uncertainty of the Plowman {\it et al.} method. However, these estimations are related to the isothermal homogeneous plasma.  Really, the coronal plasma is multi-temperature and structured, when different components overlap along the line-of-sight, so these values represent only the minimal error estimations.

  \begin{figure}    
   \centerline{\includegraphics[width=1.05\textwidth,clip=]{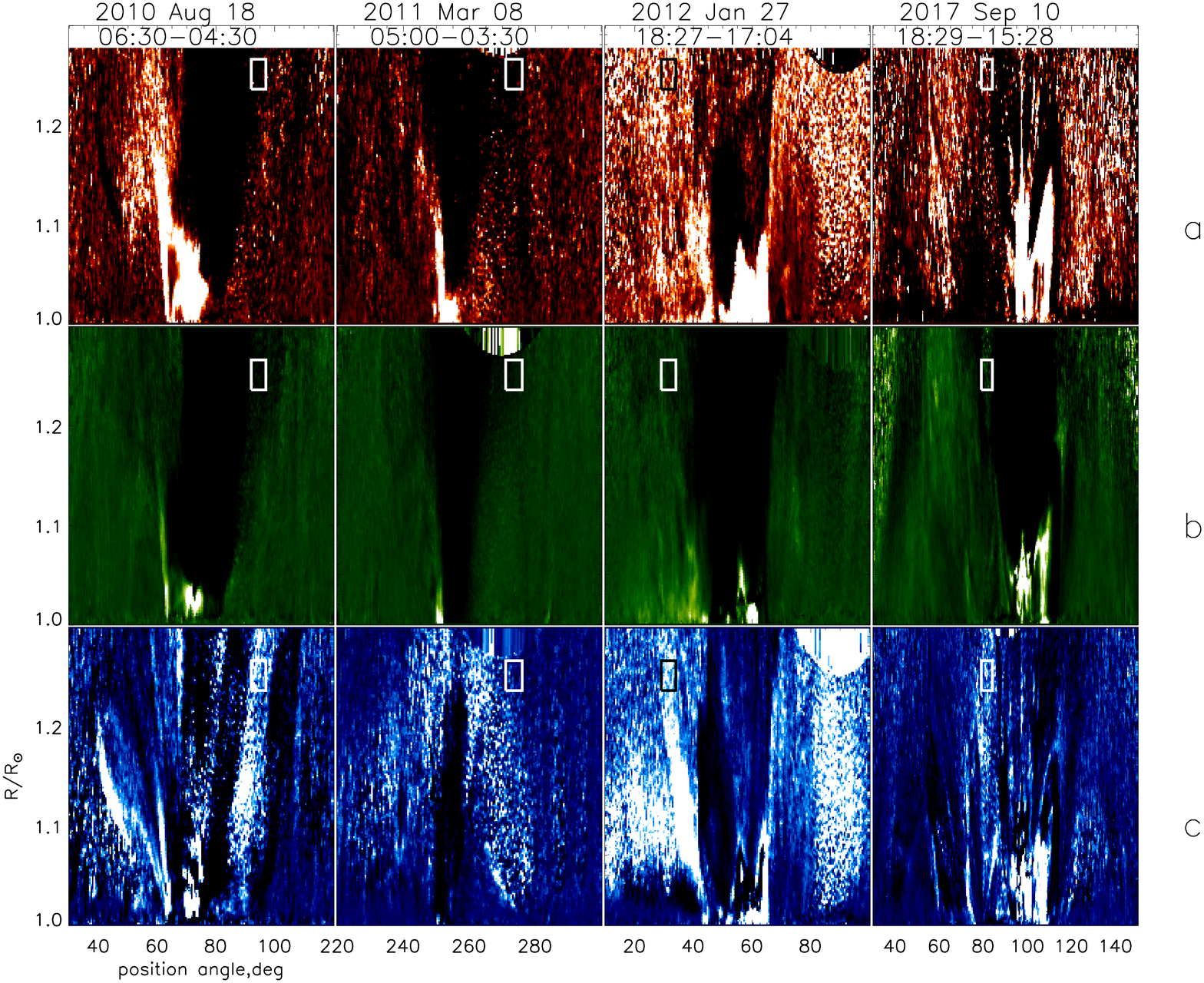}
              }
              \caption{Relative-difference EM maps for the dimming regions for the events under study in three temperature ranges: 0.3\,--\,1.3\,MK (blue), 1.3\,--\,3.7\,MK (green), and 3.7\,--\,30\,MK (red). Rectangles with white contours indicate the sites used for the DEM analysis (see Figure~\ref{F-8} below). In the titles of each column, the dimming and reference times are shown similar to Figure~\ref{F-2}.
                      }
   \label{F-7}
   \end{figure}

  \begin{figure}    
   \centerline{
               \includegraphics[width=1.05\textwidth,clip=]{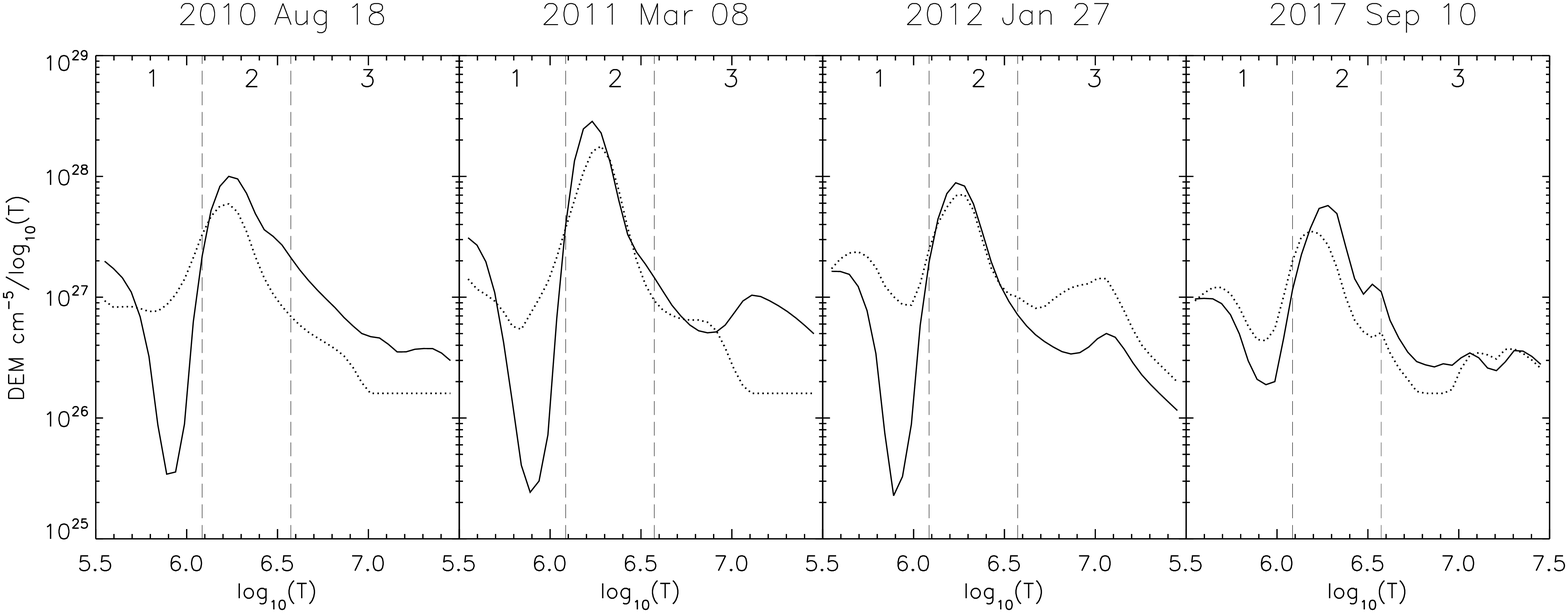}
              }
\caption{DEM temperature distributions for the regions of the cold ray-like features in the dimmings shown in Figures~\ref{F-2} and \ref{F-7} (times listed in Table~\ref{T-2}): solid lines correspond to the reference times (before eruption, first value in Table~\ref{T-2}), the dotted lines stand for the times near the dimming maximum (the second value in Table~\ref{T-2}). The temperature intervals 0.3\,--\,1.3, 1.3\,--\,3.7, and 3.7\,--\,30\,MK are marked with numbers 1, 2, and 3.
        }
   \label{F-8}
   \end{figure}

Figure~\ref{F-7} shows the colored maps of the relative difference EM values [$(\mathrm{EM}(t)-\mathrm{EM}(t_{\mathrm{ref}}))/\mathrm{EM}(t_{\mathrm{ref}})$] for the dimming regions of the events under study in three temperature ranges. Rectangles on the DEM maps indicated in Figure~\ref{F-7} correspond to the regions of the ray-like features in the SWAP maps shown in Figure~\ref{F-2}. Figure~\ref{F-8} displays the DEM temperature distributions averaged over the selected regions in the SWAP rays before and after eruption. It is seen that the DEM values noticeably increased for the cold plasma component in the rays compared to the pre-eruptive state at the temperatures of $\log_{10}(T[K])\approx 5.8-5.9$ ($T\approx 0.6-0.8$~MK).

Table~\ref{T-2} shows the EM values integrated over each of the temperature ranges and averaged over the marked regions of the rays and the values of EM-weighted temperature evaluated for the times before eruption and for the dimming maximum. As seen in Table~\ref{T-2}, the EM values in the rays at the dimming state dropped in II interval (the medium temperature peak) by 19\,--\,43\,\%, and in III interval (hot peak) by 26\,--\,47\,\%. In contrast, EM in the colder I interval was noticeably enhanced up to a factor of two in the dimming in comparison with the pre-eruption state (except the event of 8 March 2011, where the growth was only 3\,\%). This result agrees with the enhanced brightness of these structures in the SWAP difference images shown in Figures~\ref{F-2} and \ref{F-3}.

\begin{table}
\caption{Variations of EM and EM-weighted temperature in the marked regions of the dimmings in the temperature ranges I (0.3\,--\,1.3\,MK), II (1.3\,--\,3.7\,MK), and III (3.7\,--\,30\,MK).
}
\label{T-2}
\begin{tabular}{rccccccccc} 
  \hline
Date & Time & &\multicolumn{3}{c}{EM  [$10^{26}$cm$^{-5}$]}&& \multicolumn{3}{c}{$\log_{10} (T[\mathrm{K}])$} \\
     &      & &  I & II & III && I & II & III  \\
  \hline
18 Aug. 2010 & 04:30 && 5.1 & 28. & 6.4 && 5.76 & 6.30  & 6.86 \\
           & 06:30 && 7.1 & 16. & 2.5 && 5.89 & 6.27  & 6.84 \\
08 Mar. 2011 & 03:30 && 7.6 & 57. & 6.9 && 5.79 & 6.26  & 6.97   \\
           & 05:00 && 7.8 & 42. & 3.7 && 5.87 & 6.29  & 6.80   \\
27 Jan. 2012 & 17:04 && 5.0 & 22. & 3.5 && 5.77 & 6.27  & 6.92  \\
           & 18:27 && 10. & 18. & 8.0 && 5.79 & 6.28  & 6.93  \\
10 Sep. 2017 & 15:28 && 3.8 & 15. & 3.5 && 5.77 & 6.30  & 6.94  \\
           & 18:29 && 5.7 & 8.8 & 2.6 && 5.82 & 6.26  & 7.03    \\
  \hline
\end{tabular}
\end{table}

It should be noted that these three temperature components reflect inhomogeneity of the plasma along the line-of-sight, which we can presumably attribute to specific structures. We suggest that the middle temperature component mostly corresponds to the ambient corona outside the dimming body. Before eruption, at the reference time it contained also plasma in the dimming volume, which disappeared during the CME formation. As is seen in Figure~\ref{F-7}, the hot component in the dimmings shows a fast decrease with height and disappears at 1.2\,$\mathrm{R}_{\odot}$; thus it most probably corresponds to the surrounding hot loops invisible in the SWAP images and does not relate to the features under study. The cold EM component evidently grows in the dimmings at the distances above 1.05\,--\,1.1\,$\mathrm{R}_{\odot}$ being especially enhanced in the ray-like features seen in the SWAP images. It means that these cold features contain plasma with the temperature corresponding to the SWAP temperature response range, which for quiet Sun extends over $T\approx 0.3 - 2$\,MK.

\section{Discussion}

\subsection{Contribution of the Cold Ray-Like Features to the Solar Wind}

\begin{figure}    
   \centerline{\includegraphics[width=1.02\textwidth,clip=]{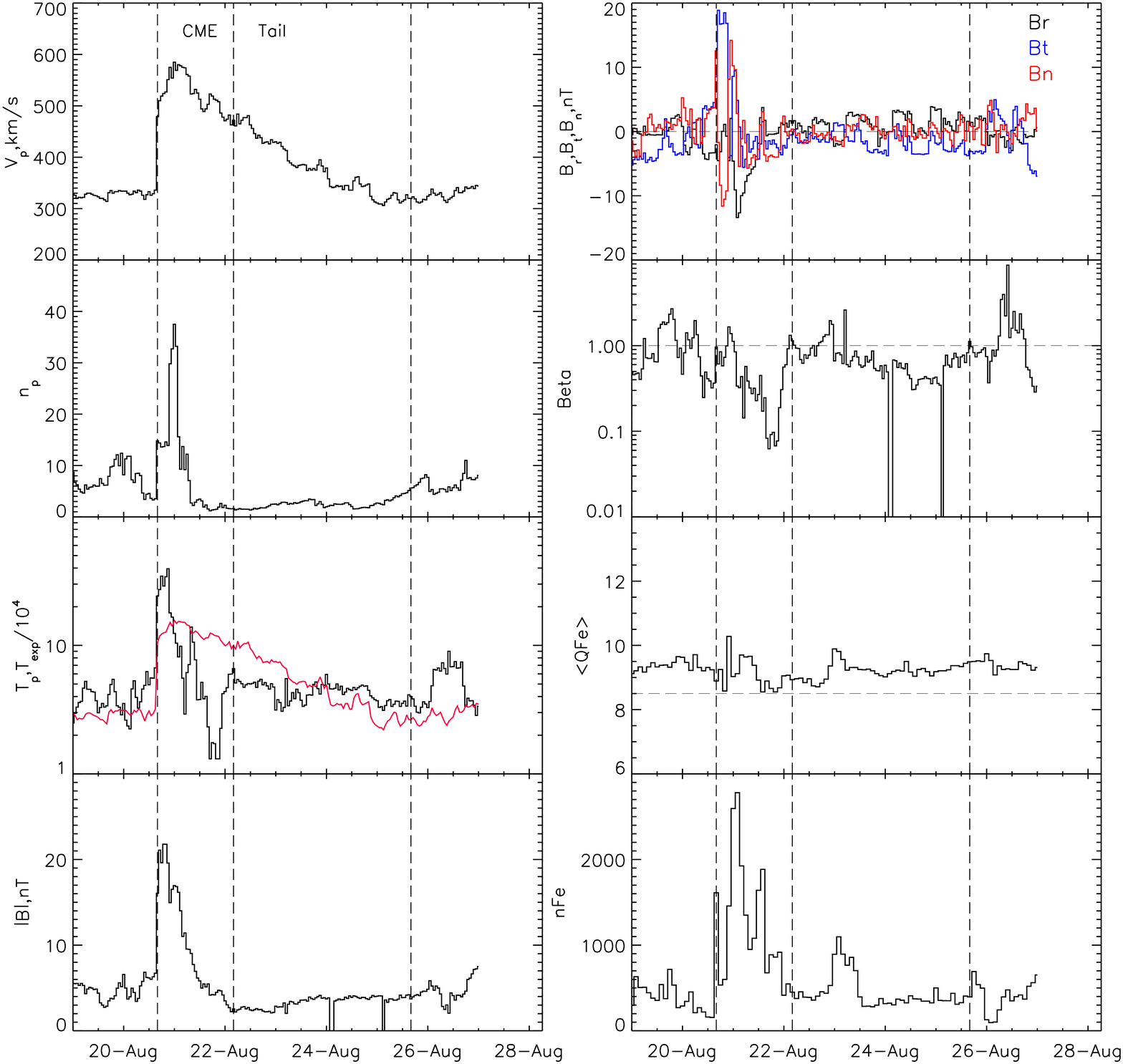}
              }
              \caption{Temporal evolution of plasma parameters in the solar wind associated with the CME event on 18 August 2010 from the STEREO-A/PLASTIC data. The red line at the temperature panel indicates the expected proton temperature $T_{\mathrm{exp}}$.
                      }
   \label{F-9}
   \end{figure}

To investigate a probable link of the ray-like features with the solar wind, we analyzed  parameters of the solar-wind transient registered by the PLASTIC instrument onboard STEREO-A (see Figure~\ref{F-9}) in connection with the first CME event in Table~\ref{T-1}, which appeared in the LASCO field of view on 18 August 2010 at 05:48:05~UT. According to the Drag-Based Model calculations (see site {\sf oh.geof.unizg.hr/DBM/dbm.php}) based on the CME data from the LASCO database (see site {\sf cdaw.gsfc.nasa.gov/CME\_list/}) the associated transient should arrive at the STEREO-A location on 20 August 2010 at 12:46~UT with the speed  of 470\,km~s$^{-1}$.

The solar-wind disturbance associated with the CME appearance was detected on 20 August 2010 at 16:00~UT by a sharp increase of the proton speed (from 336 to 479\,km~s$^{-1}$), density, and magnetic-field magnitude. The magnetic-field disturbance, seen as a sharp increase of the strength and variation of the magnetic field components (in the Radial Tangential Normal (RTN) coordinate system), lasted up to 22 August (02:00~UT), which may be identified as a passing of the CME front followed by a magnetic cloud. The rise of the proton temperature and the rise of the derived $\beta$-parameter to unity also confirm the end of the CME at that moment. Only the proton speed remained at the enhanced level of 500\,km~s$^{-1}$ and decreased to the level of slow solar wind (340\,km~s$^{-1}$) on 24 August (12:00~UT). We suppose that in the period between the end of the magnetic cloud and decrease of the proton speed to the slow solar-wind level (the CME tail) the solar wind contained plasma from the stream seen in the dimming by SWAP as cold ray on 18 August between 05:00 and 08:00~UT.

We assume that the plasma seen in the cold ray was propagating in the heliosphere behind the CME body along open magnetic-field lines with a constant speed equal to the solar-wind speed at the beginning of the CME tail (a drag decelerates only the CME front). To the distance of 20\,$\mathrm{R}_{\odot}$ (as it is ordinarily considered in the CME propagation models), the stream of cold plasma moved during 15 hours with a constant acceleration and the mean speed of 250\,km~s$^{-1}$; then the stream propagated in the heliosphere during 75 hours to the STEREO-A spacecraft with the speed of 500\,km~s$^{-1}$. Thus, the stream would arrive on 21 August at about 23:00~UT, which is very close to the beginning of the CME tail (22 August, 04:00~UT), according to the measurements.

The arrival of the cold stream is seen by the $\beta$-parameter increase to unity and by enhancement of the mean charge of Fe-ions to $Q_{\mathrm{Fe}}=10$ and the increase of the total Fe-ion flux. This Fe mean charge corresponds to the plasma temperature of $\approx$1\,MK measured in the cold rays at the heights of 1.25\,$\mathrm{R}_{\odot}$. In analogy with the fast and slow solar-wind models (see, {\it e.g.}, \citealp{Landi12}), the mean charge $Q_{\mathrm{Fe}}=10$ of the stream  practically does not change up to the freezing-in distances $\approx 3-4\,\mathrm{R}_{\odot}$ and remains unchanged later in the heliosphere. The transient solar wind stream arising in the dimming and appearing in the CME tail has signatures of the slow solar wind, except for the enhanced speed. This result explains the observations of some CMEs by \cite{Temmer17A}, who found prolonged (up to six days after arrival of a CME) excess of the solar-wind speed above the background.

\subsection{Comparison of Cold Ray-Like Features Seen by SWAP with the Rays Observed in White Light/UV by LASCO and UVCS}

\cite{Webb03}, \cite{Bemporad06}, \cite{Vrsnak09}, and \cite{Ciaravella08} pointed out that the hot rays with temperatures of 3\,--\,8\,MK appearing in the outer corona in the aftermath of CMEs represent heliospheric current sheets connecting the post-eruptive arcades in the eruption site to the CME front. \cite{Ciaravella13} defined three types of the rays appearing during the CME events by their emission in the spectral lines of different temperatures: the hot rays seen in Fe~{\sc xviii} ion lines, the coronal rays detected in those of Si~{\sc xii} and O~{\sc vi}, and the cool rays that show emission only in the lines of O~{\sc vi} and/or $\mathrm{Ly}\alpha$.

  \begin{figure}    
   \centerline{\includegraphics[width=1.05\textwidth,clip=]{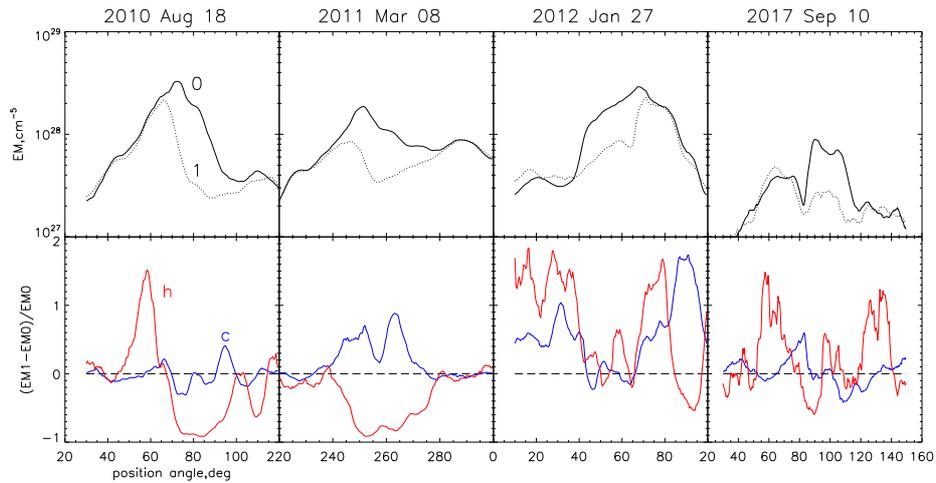}
              }
              \caption{Top panels: the averaged absolute EM values in the ray-like features  integrated over the full temperature range 0.3\,--\,30\,MK for the event on 18 August 2010. The solid lines are the initial EM values before eruption, the dotted lines are those in the dimmings (see Table~\ref{T-2}). Bottom panels: the blue lines (marked as ``c'') stand for the relative difference EM values of the cold component (0.3\,--\,1.3\,MK), the red lines (marked as ``h'') correspond to those of the hot component (3.7\,--\,30\,MK).
                      }
   \label{F-10}
   \end{figure}

They identified the coronal rays with $T=1.5-3$\,MK as corresponding to the ambient plasma, and the cool rays with $T < 1.5-2$\,MK as associated with the prominence material falling down on the Sun. To define the signatures of these rays  in the lower corona, we made a scan of the plasma properties derived from the DEM distributions over the dimming areas at the distance of 1.25\,$\mathrm{R}_{\odot}$ (see Figure~\ref{F-10}). At the upper panel there are latitudinal profiles of EM at the distance of $1.25\pm 0.07\,\mathrm{R}_{\odot}$ obtained from the DEM maps integrated over the full temperature interval (0.3\,--\,30\,MK): the solid lines correspond to the initial (before eruption) states, while the dotted ones correspond to the dimming states. The times of these states are given in Table~\ref{T-2}. At the lower panel there are the relative difference EM distributions (EM1 - EM0)/EM0 (EM0 and EM1 are the initial and post-eruptive EM values respectively). These distributions are given for the cold (the blue lines) and hot (the red lines) components of the plasma in accordance with the temperature ranges I and III in Table~\ref{T-2}.

\begin{table}
\caption{The EM-weighted temperatures for cold and hot ray-like features at $R=1.25\,\mathrm{R}_{\odot}$ averaged over the angular ranges for each event indicated in Figure~\ref{F-10}.
}
\label{T-3}
\begin{tabular}{lccc}     
  \hline                   
 Event & Time , & $\langle T_{\mathrm{cold}}\rangle_{\mathrm{EM}}$,   & $\langle T_{\mathrm{hot}}\rangle_{\mathrm{EM}}$,   \\
       &  [UT]    &    [MK]     & [MK]     \\
  \hline
  18 Aug. 2010 &  06:30  & 0.82$\pm$0.26\tabnote{The errors include the standard deviation of averaging the temperatures (0.07\,--\,0.08\,MK) and uncertainty of the DEM method at the temperatures $T\approx 1-10$\,MK (0.26\,MK).} &  8.4$\pm$2.2  \\
 08 Mar. 2011 &  05:00  & 0.74$\pm$0.26 &  8.6$\pm$1.9  \\
 27 Jan. 2012 &  18:27  & 0.63$\pm$0.27 &  9.8$\pm$1.0  \\
 10 Sep. 2017 &  18:29  & 0.65$\pm$0.26 &  13.1$\pm$1.5  \\
  \hline
\end{tabular}
\end{table}

These data show that the regions of the cold rays seen by SWAP are located mostly in the dimming areas, but in the two last cases they are also seen outside. The hottest structures are located mostly aside of dimmings and are not seen in the SWAP spectral passband because of their high temperature conditions. Table~\ref{T-3} contains the temperatures of the cold and hot rays obtained from the DEM distributions by the EM-weighting and then averaged over position angle. The cold-plasma components in all cases have practically the same temperatures and agree with the measured mean Fe-ion charge. In contrast, the temperatures of the hot components are more variable and obviously do not reproduce the corresponding ion component of solar wind with $Q_{\mathrm{Fe}} > 12$, as follows from the ionization balance equations (\citealp{Rodkin17}; \citealp{Grechnev19}). Taking into account their location outside dimmings, we can conclude that in our cases the hot components did not contribute to the solar wind but probably were related to the hot post-eruptive loops. Table~\ref{T-4} summarizes properties of the cold ray-like features obtained from the SWAP data in the events under study compared to the properties of the rays identified by \cite{Ciaravella13} from an analysis of the UVCS spectrograms and LASCO data.

\begin{table}
\caption{Parameters of the ray-like features observed by SWAP and the rays observed by LASCO and UVCS (\citealp{Ciaravella13}).
}
\label{T-4}
\begin{tabular}{lcccc}     
  \hline                   
 Ray-like feature    & Cold rays  & Hot rays

  & Coronal rays  & Cool rays   \\
   type   & (SWAP) & (Fe~{\sc xviii})  & (Si~{\sc xii}, O~{\sc vi}) & (LASCO \& UVCS)      \\
  \hline
 Observational region  &  Dimming   & Corona   & Corona  & Corona   \\
  & $1.1-1.7\,\mathrm{R}_{\odot}$ & $ > 1.5 \,\mathrm{R}_{\odot}$ & $1.4-2.3\,\mathrm{R}_{\odot}$ & $ 1.5 - 4.1\,\mathrm{R}_{\odot}$ \\
 Delay of appearance  & $<$~1 hr & 1\,--\,4 hrs & 1\,--\,30 hrs & 0.5\,--\,17 hrs \\
 after start of a CME  &  &   &   &    \\
 Duration &  40 min\,--\,16 hrs & 2\,--\,38 hrs & 5 hrs\,--\,3 days & 3\,--\,24 hrs \\
 Temperature &  $<$1.5\,MK & 3\,--\,8\,MK & 1.5\,--\,3\,MK & $<$~1.5\,MK  \\
  \hline
\end{tabular}
\end{table}

It follows that the ray-like features observed by SWAP are similar to those classified by \cite{Ciaravella13} as cool rays. However, there are several differences. The ray-like features in the SWAP images appeared in less than one hour after the CME onset and existed during the dimming life span, whereas LASCO typically observed the coronal and cool rays one hour later the CME onsets during many hours up to several days not related to the dimming existence. We suppose that SWAP observed origins of the cool ray-like features at their early phase of formation in the dimming regions.

\subsection{Nature of the Cool Ray-Like Features}

\begin{figure}    
   \centerline{\includegraphics[width=0.7\textwidth,clip=]{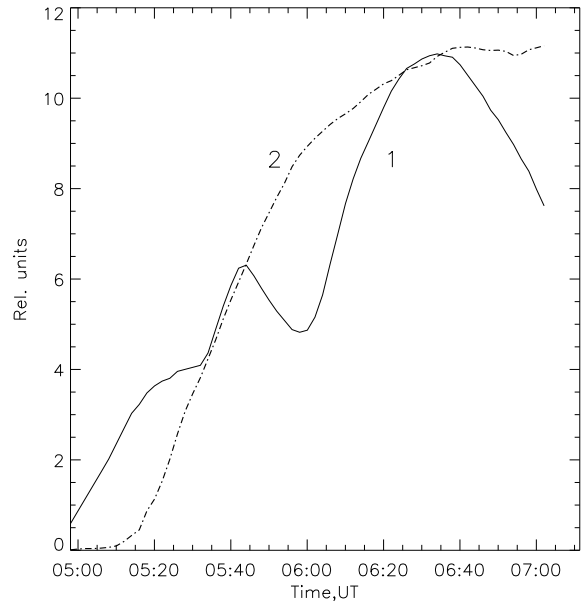}
              }
              \caption{Temporal variation of brightness (value $-y(t)$ from Eq.~(\ref{Eq-1})) in the cold ray observed by SWAP on 18 August 2010 (curve 1) in comparison with the integrated intensity of the dimming in the SDO/AIA 211\,\AA\ band taken from the Solar Demon data (curve 2).
                      }
   \label{F-11}
   \end{figure}

In Figure~\ref{F-11} we represent the time evolution of the value $-y(t)$ (see Eq.~(\ref{Eq-1})) for the region marked in Figure~\ref{F-7} on the event of 18 August 2010 as compared to the same from the Solar Demon data. Figure~\ref{F-11} shows that temporal variation of the intensity for the event of 18 August 2010 is well correlated with the total decrease of the EUV emission in the dimming taken from the Solar Demon data. The Pearson correlation coefficient between these functions is about 0.9. It suggests that the cold rays observed by SWAP contain matter released from the dimming volume, probably by reconnection between the quasi-open magnetic-field lines of the expanding flux rope and cool magnetic loops surrounding the active region. Several maxima at the ray light curve may be presumably caused by the moving blobs.

The supposition of \cite{Ciaravella13} that cool rays are formed from the cool matter of a prominence body falling down on the Sun from a CME seems less plausible for two reasons: first, as it follows from \cite{Tripathi07} the down-falling parts of prominences do not take the form of rays. Second, in this case appearance and brightness of the rays would not be correlated with the dimming formation.

\begin{figure}    
   \centerline{\includegraphics[width=1.05\textwidth,clip=]{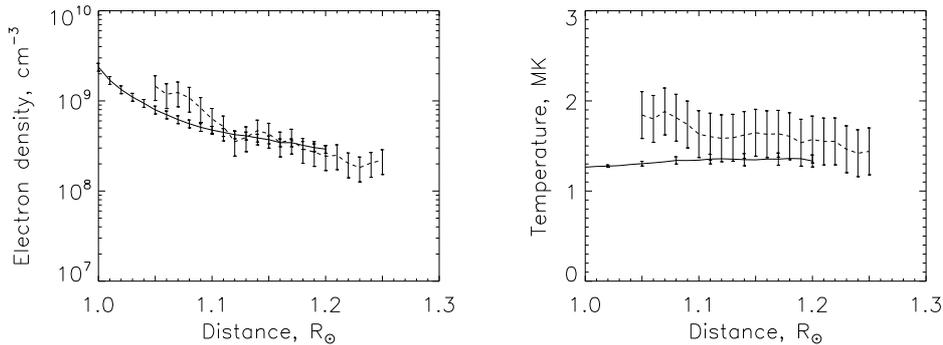}
              }
              \caption{Variation of the plasma density and temperature in the marked region of the cold ray (on 18 August 2010, 06:25~UT) in comparison with those in the streamer observed on 20 October 2010 by \cite{Goryaev14}. The dashed lines with error bars correspond to the data for the ray, and the solid lines stand for the data for the streamer.
                      }
   \label{F-12}
   \end{figure}

Figure~\ref{F-12} presents the results of analysis of the plasma density and temperature distributions along the cold ray derived from the DEM map on 18 August 2010 in comparison with those distributions for a streamer observed on 20 October 2010 (\citealp{Goryaev14}). Because in the streamer case the medium-temperature component of the ambient plasma was not subtracted, in the case of the cold ray we integrated EM over the extended temperature range 0.3\,--\,3.7\,MK.

It appears that the density and temperature height distributions in the cold ray observed by SWAP at the distances $1.1-1.25\,\mathrm{R}_{\odot}$ are close to that observed in the streamer studied at similar solar activity. A slight enhancement at the distance below 1.1\,$\mathrm{R}_{\odot}$ can be explained by overlapping with the lower post-eruptive loops. This result suggests that in both cases formation of the plasma streams contributing to solar wind is physically similar.

\section{Summary and Conclusions}

We studied four coronal dimmings associated with limb CMEs that occurred in the period 2010\,--\,2017 using PROBA2/SWAP and SDO/AIA data. Below we summarize our main results and conclusions:

\smallskip

\noindent i) Bright ray-like structures revealed in the dimming regions by SWAP at the distances up to 1.6\,$\mathrm{R}_{\odot}$ are very likely bounded by the expanding flux ropes, probably representing their trunks or lateral flows of plasma formed as a result of reconnection of neighboring magnetic loops with ``open'' magnetic-field lines of the flux rope.

\smallskip

\noindent ii) In the low corona, these structures presumably originated as fan rays associated with open magnetic structures. Such structures often appear in the origins of pseudo-streamers or in the adjacent regions between active regions and coronal holes.

\smallskip

\noindent iii) The cold rays contain plasma with a temperature below 1\,MK. It was found that the radial distributions of the electron density and temperature in the cold rays are similar to that determined by \cite{Goryaev14} in the unipolar pseudo-streamer with signatures of the plasma outflows. Temporal variations of coronal-ray brightness may be caused by the blob structure of the moving plasma.

\smallskip

\noindent iv) Due to its low temperature, the plasma in the rays is enriched by the Fe$^{8+}$--Fe$^{10+}$ ions. For the event on 18 August 2010, enhancements of densities of these ions were observed in the STEREO/PLASTIC solar-wind data.

\smallskip

\noindent v) To summarize, the ray-like structures observed by SWAP presumably represent transient sources of plasma flows with the signatures of slow solar wind (except for increased speed), propagating in the tails of CMEs. Observations of the eruption in the aftermath of the CME in the corona with the wide-field SWAP telescope in the 174\,\AA\ band allow identification of origins of these rays, which is important for forecasting of solar wind.

%

%
\begin{acks}

This study was supported by the Russian Science Foundation (RSF) under grant 17-12-01567.
We appreciate the NASA/SDO and the AIA science team; the NASA's STEREO/
PLASTIC science and instrument teams; the teams operating LASCO on SOHO.
SWAP is a project of the Centre Spatial de Li\`ege and the Royal Observatory
of Belgium funded by the Belgian Federal Science Policy Office (BELSPO). The STEREO/SECCHI data are produced by an international consortium of the Naval Research Laboratory (USA), Lockheed Martin Solar and Astrophysics Lab (USA), NASA Goddard Space Flight Center (USA), Rutherford Appleton Laboratory (UK), University of Birmingham (UK), Max-Planck-Institut f\"ur Sonnenforschung (Germany), Centre Spatiale de Li\`ege (Belgium), Institut d'Optique Th\'eorique et Appliqu\'ee (France), and Institut d'Astrophysique Spatiale (France).

\end{acks}

\medskip

\noindent {\bf Disclosure of Potential Conflict of Interest} The authors declare that they have no conflict of interest.

%
%
%


 \bibliographystyle{spr-mp-sola}
 \bibliography{sola_bibliography}

%
%
%
%

\end{article}
\end{document}